\documentclass{article}

\usepackage{amssymb,amsfonts,amsmath,stmaryrd}
\usepackage{cite,enumerate,float,indentfirst}
\usepackage{color}
\usepackage{tikz}
\usetikzlibrary{arrows,backgrounds}
\usepackage{url}
\usepackage{hyperref}
\usepackage{textcomp}

\def\be{\begin{eqnarray}}
\def\ee{\end{eqnarray}}

\def\p{\partial}
\def\tr{{\rm tr}\,}

\def\Hur{{\rm Hur}\,}
\def\Pf{{\rm Pf}\,}
\def\SP{{\rm SP}\,}
\def\OP{{\rm OP}\,}
\def\spin{{\rm spin}}

\definecolor{red}{rgb}{1,0,0}
\definecolor{orange}{rgb}{1,0.5,0}
\definecolor{violet}{rgb}{0.7,0,1}

\hoffset= - 1.0in         % switch off for draft style
\begin{document}

%???to the memory of S.Natanzon

\title{\vspace{1cm}{\Large {\bf
Connection between cut-and-join and Casimir operators
    }
    \date{}
    \author{
			{\bf A. Mironov$^{a,b,c}$}\thanks{mironov@lpi.ru; mironov@itep.ru},
			{\bf A. Morozov$^{d,b,c}$}\thanks{morozov@itep.ru},
			{\bf A. Zhabin$^{d,b}$}\thanks{alexander.zhabin@yandex.ru}
			}
}}

\maketitle
\vspace{-5.5cm}

\begin{center}
    \hfill FIAN/TH-06/21 \\
	\hfill ITEP/TH-12/21 \\
	\hfill IITP/TH-09/21 \\
	\hfill MIPT/TH-08/21
\end{center}

\vspace{3cm}

\begin{center}
   $^a$ {\small {\it Lebedev Physics Institute, Moscow 119991, Russia}}\\
   $^b$ {\small {\it Institute for Theoretical and Experimental Physics, Moscow 117218, Russia}}\\
   $^c$ {\small {\it Institute for Information Transmission Problems, Moscow 127994, Russia}}\\
   $^d$ {\small {\it Moscow Institute of Physics and Technology, Dolgoprudny 141701, Russia }} \\
   \vspace{0.25cm}

\end{center}

\vspace{0.5cm}

\begin{abstract}
We study cut-and-join operators for spin Hurwitz partition functions. We provide explicit expressions for these operators in terms of derivatives in $p$-variables without straightforward matrix realization, which is yet to be found. With the help of these expressions spin cut-and-join operators can be calculated directly and algorithmically. The reason why it is possible is the connection between mentioned operators and specially chosen Casimir operators that are easy to compute. An essential part of the connection involves shifted Q-Schur functions.
\end{abstract}

\bigskip

\section{Introduction}\label{sec:introduction}

Hurwitz numbers are well studied combinatorial objects counting ramified coverings of a Riemann surface. Generating functions for the Hurwitz numbers have found various applications in theoretical and mathematical physics as partition functions of various topological theories \cite{dijkgraaf1995mirror,dijkgraaf1989geometrical}. The most important partition functions are known to be integrable, i.e. they are $\tau$-functions of some integrable hierarchy. The simplest non-trivial example is the generating function for simple Hurwitz numbers which is a partition function of 2d Yang-Mills theory \cite{rusakov1990loop} and, moreover, is a KP $\tau$-function \cite{GKM2,okounkov2000toda,okounkov2002gromov,Lando,mironov2011complete,AMMN}. Besides, many Hurwitz partition functions may be represented as matrix models \cite{Kaz,KonHurMM,Morozov_2009,alexandrov2011matrix} and can be determined with the help of topological recursion \cite{BM,KonHurMM,borot2011matrix,dunin2019loop}.

Another natural way to study Hurwitz partition functions uses the approach of cut-and-join operators \cite{GJ,mironov2011complete,MMN2}. These operators act on very simple objects and allow one to generate a huge set of Hurwitz partition functions. Action of such operators is similar to interpolating in the space of partition functions. There are several equivalent definitions of cut-and-join operators, the most explicit one being
\begin{equation}\label{W_matrix_form}
    \hat{W}_{\Delta} = \frac{1}{z_{\Delta}} :\prod_{i=1}^{l(\Delta)} \tr \hat D^{\Delta_{i}} :
\end{equation}
where $\Delta_{i}$ are parts of the partition $\Delta = \{\Delta_1 \geq \Delta_2 \geq \ldots \Delta_{l(\Delta)}>0\}$, $z_\Delta$ is the standard symmetric factor of the Young diagram,
and the matrix operator $\hat D_{ab} := \sum_{c=1}^N X_{ac}\frac{\p}{\p X_{bc}}$ so that the normal ordering $:\ldots :$ means all the derivatives placed to the right. The Schur polynomials \cite{Mac,Fulton} are eigenfunctions of the cut-and-join operators
\begin{equation}\label{W_def_intro}
    \hat{W}_{\Delta} S_{R} = \phi_{R}(\Delta) S_{R}
\end{equation}
with eigenvalues equal to properly normalized symmetric group characters (see below).

Theory of the ordinary Hurwitz numbers as described above is well-known. However, theory of the \textit{spin} Hurwitz numbers counting coverings with spin structure lacks enough information: a counterpart of the explicit expression \eqref{W_matrix_form} is yet unknown. It is still possible to study the cut-and-join operators in the spin case taking the counterpart of \eqref{W_def_intro} as a definition:
\begin{equation}\label{spin_W_def_intro}
    \hat{\mathcal{W}}_{\Delta} Q_{R} = \Phi_{R}(\Delta) Q_{R}
\end{equation}
where $Q_{R}$ are the Q-Schur polynomials \cite{Schur,Mac}, and $\Phi_{R}(\Delta)$ are properly normalized characters of the Sergeev group \cite{Serg}. However, such a definition does not provide an explicit expression for the spin cut-and-join operators in a matrix form or in terms of derivatives.

The aim of this paper is to construct an algorithmic way to obtain explicit expressions for the spin cut-and-join operators. Our main result is formula (\ref{QW_via_Casimirs_spin}), which establishes a connection between eigenvalues of the cut-and-join operators and specially chosen Casimir operators. These Casimir operators can be constructed using the approach of \cite{mironov2020around}. This allows us to evaluate immediately cut-and-join operators, while in \cite{mironov2020cut} they were obtained by solving a system of equations in the space of operators, which being a laborious procedure allowed one to evaluate in practice only a few first operators.

Such a connection exists both for the ordinary and spin Hurwitz numbers. Therefore, one may express the cut-and-join operators via the Casimir operators. In the spin case, the corresponding Casimir operators can be easily calculated as zero modes of the $W_{1+\infty}^{B}$ algebra of symmetries of the BKP integrable hierarchy.

The paper is organized as follows. We review all the necessary information about the ordinary $\hat{W}$-operators in section 2. Then we generalize all the formulas to the spin case in section 3. Section 4 is devoted to an explicit calculation of the Casimir operators via bosonization formulas, and then we provide expressions for the spin cut-and-join operators. In Appendix A, we compare different normalization factors in definitions of the spin Hurwitz numbers in various papers, and provide first non-trivial examples of these numbers.

\section{Ordinary Hurwitz numbers}

\subsection{Basic objects}

First of all, we fix the notation:
\begin{itemize}
    \item Let $R$ denotes a partition (Young diagram) $ R = \{ R_{1} \ge R_{2} \ge \dots R_{l(R)} > 0 \}$ with the length $l(R)$, and $|R|$ is a sum of all $R_{i}$. Another way to denote the partition is $[1^{m_{1}}, 2^{m_{2}}, 3^{m_{3}}, \dots]$, where $m_{k}$ is the number of lines of length $k$.

    \item $S_{R}(p)$ is the Schur polynomial in $p$-variables ($p_{k} = \sum x_{i}^{k}$),

    $\psi_{R}(\Delta)$ is the character of symmetric group defined for $|R| = |\Delta|$,

    $z_{\Delta} := \prod_{k\ge1} k^{m_{k}} m_{k}!$ is the order of automorphism of the partition $\Delta$.

    All these notions come together in one Fr\"obenius formula for the Schur polynomials:
    \begin{equation}\label{Schur_via_characters}
        S_{R}(p) = \sum_{\Delta \vdash |R|} \frac{\psi_{R}(\Delta)}{z_{\Delta}} p_{\Delta}
    \end{equation}
    where $p_{\Delta}:= \prod_{i=1}^{l(\Delta)} p_{\Delta_{i}}$. This simple formula reveals the Schur-Weyl duality between the general linear and symmetric groups.

    \item $d_{R} := S_{R}(\delta_{k,1}) = \frac{\psi_{R}([1^{|R|}])}{|R|!}$ is the normalized dimension of an irreducible representation $R$ of the symmetric group ${\cal S}_{|R|}$,

    $\phi_{R}(\Delta) := \frac{\psi_{R}(\Delta)}{z_{\Delta} d_{R}}$ is the normalized characters of symmetric group. Here, it is still defined only for $|R|=|\Delta|$. We provide a definition of $\phi_{R}(\Delta)$ for arbitrary $R$ and $\Delta$ later.

    Formula \eqref{Schur_via_characters} can be rewritten in the form
    \begin{equation}\label{Schur_via_normalized_characters}
        S_{R}(p) = \sum_{\Delta \vdash |R|} d_{R} \phi_{R}(\Delta) p_{\Delta}
    \end{equation}
\end{itemize}

The ordinary Hurwitz numbers count the $n$-sheet ramified coverings of Riemann surface of genus $g$ with $m$ ramification points \cite{Hur,Burn}. The type of ramification at the $i$-th point is given by the partition $\Delta_{i}$, $i \in \overline{1,m}$. We use the following Fr\"obenius formula as the definition of ordinary Hurwitz numbers \cite{Fro}:
\begin{equation}\label{Hurwitz_definition}
    \Hur_{g,n} (\Delta_{1}, \Delta_{2}, \ldots, \Delta_{m}) := \sum_{R \vdash n} d_{R}^{2-2g} \phi_{R}(\Delta_{1}) \phi_{R}(\Delta_{2}) \ldots \phi_{R}(\Delta_{m})
\end{equation}
where $|\Delta_{1}| = |\Delta_{2}| = \dots = |\Delta_{m}| = n$. If we fix the number of ramification points, the Hurwitz numbers can be organized in a simple generating function:
\begin{equation}\label{Hurwitz_partition_function_easy}
    \sum_{n} \sum_{\Delta_{1}, \dots, \Delta_{m}} \Hur_{g,n}(\Delta_{1}, \dots, \Delta_{m}) p_{\Delta_{1}}^{(1)} \dots p_{\Delta_{m}}^{(m)} = \sum_{R} d_{R}^{2-2g-m} S_{R}(p^{(1)}) \dots S_{R}(p^{(m)})
\end{equation}
Here the sum in the r.h.s. is taken over all partitions $R$. One may obtain such a generating function using \eqref{Schur_via_normalized_characters} and \eqref{Hurwitz_definition}.

Other generating functions are restricted not by the number of ramification points, but rather by type of ramification. The most general Hurwitz partition function is
\begin{equation}\label{Hurwitz_partition_function_general}
    Z_{g}(p^{(1)}, \dots, p^{(m)}| \beta) = \sum_{R} \left( e^{\sum_{\Delta} \beta_{\Delta} \phi_{R}(\Delta)} \right) d_{R}^{2-2g-m} S_{R}(p^{(1)}) \dots S_{R}(p^{(m)})
\end{equation}
where $\beta_{\Delta}$ are arbitrary parameters, and the definition of $\phi_{R}(\Delta)$ is extended to arbitrary $R$ and $\Delta$ as follows \cite{IK,mironov2011complete,MMN2}:
\begin{equation}\label{phi_normalized_extended}
    \phi_{R}(\Delta) := \begin{cases}
    0, & |\Delta| > |R| \\
    \frac{(|R| - |\Delta| + r)!}{r! (|R| - |\Delta|)!} \phi_{R}([\Delta, 1^{|R|-|\Delta|}]), & |\Delta| \le |R|
    \end{cases}
\end{equation}
$r$ is the number of lines of unit length in $\Delta$. For $|\Delta| = |R|$, such a definition is precisely the normalized symmetric group characters. Note that the generating function \eqref{Hurwitz_partition_function_easy} is obtained from \eqref{Hurwitz_partition_function_general} when all $\beta_{\Delta}=0$.

Well-known examples of the partition function \eqref{Hurwitz_partition_function_general} are the simple and double simple Hurwitz numbers counting ramifications of Riemann sphere ($g=0$). The generating function for the simple Hurwitz numbers is
\begin{equation}
    Z(p) = \sum_{R} e^{u \phi_{R}([2])} d_{R} S_{R}(p)
\end{equation}
with the parameters $\beta_{[2]} = u, \; \beta_{\Delta} = 0 \; (\Delta \ne [2])$. The $m=1$ case corresponds to one point with an arbitrary ramification, the other ramification points being simple. The case of $m=2$ reproduces the generating function for the double simple Hurwitz numbers:
\begin{equation}
    Z(p, \bar{p}) = \sum_{R} e^{u \phi_{R}([2])} S_{R}(p) S_{R}(\bar{p})
\end{equation}
Both these examples are $\tau$-functions of the KP/Toda hierarchy. Integrability of the general Hurwitz partition function \eqref{Hurwitz_partition_function_general} depends on the choice of parameters $\beta_{\Delta}$ so that when the symmetric group characters combine into linear combination of certain Casimir eigenvalues (defined in \eqref{gl_Casimirs}), \eqref{Hurwitz_partition_function_general} becomes a $\tau$-function.

\subsection{Shifted Schur polynomials}

In order to move further, we need to introduce a notion of shifted Schur polynomials.
In calculation of the shifted Schur polynomials, we follow the original work of \cite{okounkov1997shifted}. Similarly to the ordinary Schur polynomials, the shifted ones are labeled by a partition $\mu$, and may be defined in symmetric function variables $x_i$ (the shifted Schur polynomials are symmetric in $x_i-i$):
\begin{equation}\label{shift_Schur_two_dets}
    S_{\mu}^{*}(x) := \frac{\det_{i,j} (x_{i} + n - i| \mu_{j} + n - j )}{\det_{i,j} (x_{i} + n - i| n-j)},
\end{equation}
where
\begin{equation}\label{shift_power_x}
    (x|n) = \begin{cases}
    1, & n=0\\
    \prod_{k=0}^{n-1} (x-k), & n\ge 1
    \end{cases}
\end{equation}
One can further change the $x$-variables to power sum $p$-variables, which, for the shifted Schur polynomials, look like
\begin{equation}
    p_{k}^{*} = \sum_{i\ge1} \Big( (x_{i}-i)^{k} - (-i)^{k} \Big)
\end{equation}
The additional shift as compared to the standard power sums is necessary to guarantee that the shifted Schur polynomials as functions of
$p^{*}_k$ do not depend on the number of variables $x_i$.

Instead of starting from polynomials of $x_i$, one can define the shifted Schur polynomials directly in the $p^{*}$-variables similar to the ordinary Schur polynomials. To this end, we start from the generating function
\begin{equation}\label{generating_function}
    H^{*}(u) := \exp \left( \sum_{k=1}^{\infty} \frac{p_{k}^{*} u^{-k}}{k} \right) = \sum_{r=0}^{\infty} \frac{h_{r}^{*}(p^{*})}{(u|r)}
\end{equation}
The first few orders of the last sum are
\begin{equation}
    H^{*}(u) = h_{0}^{*} + \frac{h_{1}^{*}}{u} + \frac{h_{2}^{*}}{u(u-1)} + \frac{h_{3}^{*}}{u(u-1)(u-2)} + \dots
\end{equation}
Expanding all terms in $(1/u)$, one obtains the expression
\begin{equation}
    H^{*}(u) = h_{0}^{*} + \frac{h_{1}^{*}}{u} + \frac{h_{2}^{*}}{u^{2}} + \frac{h_{3}^{*} + h_{2}^{*}}{u^{3}} + \dots
\end{equation}
from which it is possible to obtain all the functions $h_{k}^{*}$ using \eqref{generating_function}. These $h_{k}^{*}$ are building blocks of the shifted Schur polynomials as well as totally symmetric functions are for the ordinary Schur polynomials.

Finally, the shifted Schur polynomials can be obtained with the help of determinant formulas. First of all, they can be expressed via $h_{k}^{*}$:
\begin{equation}\label{shift_Schur_det}
    S_{\mu}^{*} = \det_{1 \le i,j \le l(\mu)} \Big[ \varphi^{j-1}(h_{\mu_{i} - i + j}^{*}) \Big],
\end{equation}
where
\begin{equation}
    \varphi^{r}(h_{k}^{*}) = \begin{cases}
    h_{0}^{*}, & k=0 \\
    \sum_{i=0}^{r} \binom{r}{i}(k-1|i) h_{k-i}^{*}, & k>0
    \end{cases}
\end{equation}
It is easy to verify that, for the partition $\mu = [n]$, the corresponding shifted Schur polynomial is precisely $h_{n}^{*}$.

Another, even more elegant determinant formula emerge in the hook (Fr\"obenius) variables parameterizing the Young diagram $\mu$ in a somewhat different way, $\mu=(\vec\alpha|\vec\beta)$ with $\alpha_1>\alpha_2>\ldots>\alpha_l>0$, $\beta_1>\beta_2>\ldots>\beta_l>0$ so that, e.g., $\mu=[3,2,1]=(3,1|3,1)$. In these variables, the shifted Schur polynomials are given \cite{MM2019hook} by
\begin{equation}
S^*_{(\vec\alpha|\vec\beta)}=\det_{i,j}S^*_{(\alpha_i|\beta_j)}
\end{equation}
where $S^*_{(\alpha_i|\beta_j)}$ is the shifted Schur polynomial associated with the one-hook Young diagram.

The first few examples of the shifted Schur polynomials are
\begin{equation}
    \begin{gathered}
    S_{[1]}^{*} = \underbrace{p_{1}^{*}}_{=S_{[1]}(p=p^{*})}\\
    S_{[2]}^{*} = \underbrace{ \frac{(p_{1}^{*})^{2}}{2} + \frac{p_{2}^{*}}{2} }_{=S_{[2]}(p=p^{*})}\\
    S_{[1,1]}^{*} = \underbrace{ \frac{(p_{1}^{*})^{2}}{2} - \frac{p_{2}^{*}}{2} }_{=S_{[1,1]}(p=p^{*})} - p_{1}^{*}\\
    S_{[3]}^{*} = \underbrace{ \frac{(p_{1}^{*})^{3}}{6} + \frac{p_{1}^{*} p_{2}^{*}}{2} + \frac{p_{3}^{*}}{3} }_{=S_{[3]}(p=p^{*})} - \frac{(p_{1}^{*})^{2}}{2} - \frac{p_{2}^{*}}{2}\\
    S_{[2,1]}^{*} = \underbrace{ \frac{(p_{1}^{*})^{3}}{3} - \frac{p_{3}^{*}}{3} }_{=S_{[2,1]}(p=p^{*})} - \frac{(p_{1}^{*})^{2}}{2} - \frac{p_{2}^{*}}{2}\\
    S_{[1,1,1]}^{*} = \underbrace{ \frac{(p_{1}^{*})^{3}}{6} - \frac{p_{1}^{*} p_{2}^{*}}{2} + \frac{p_{3}^{*}}{3} }_{=S_{[1,1,1]}(p=p^{*})} - \frac{3(p_{1}^{*})^{2}}{2} + \frac{3p_{2}^{*}}{2} - 2p_{1}^{*}
    \end{gathered}
\end{equation}
The shifted Schur polynomials may be also evaluated via the ordinary ones using the lower-triangular expansion
\begin{equation}\label{shifted_Schur_property}
    S_{\mu}^{*}(p^{*}) = S_{\mu}(p^{*}) + \sum_{\lambda: \; |\lambda| < |\mu|} c_{\mu,\lambda} S_{\lambda}(p^{*})
\end{equation}
with coefficients $c_{\lambda,\mu}$ uniquely determined from the system of equations
\begin{equation}\label{shifted_system}
    S_{\mu}^{*}(x_{i} = R_{i}) = 0, \;\;\; \mu \nsubseteq R
\end{equation}
Such a property is remarkable for several reasons. First of all, the uniqueness of coefficients $c_{\mu,\lambda}$ allows one to use it as a definition of the shifted Schur functions. Secondly, it is probably the most convenient way to obtain the shifted Schur polynomials in the $p^{*}$ variables. Thirdly, this property holds for the generalization to the shifted Q-Schur polynomials.

\subsection{Ordinary cut-and-join operators}

Since the generating functions of type \eqref{Hurwitz_partition_function_easy} and \eqref{Hurwitz_partition_function_general} look very similar, it is natural to define operators that determine interpolation between them in the space of parameters $\beta$. These operators are called cut-and-join operators and may be defined as \cite{mironov2011complete}
\begin{equation}\label{W_definition}
    \hat{W}_{\Delta} S_{R}(p) := \phi_{R}(\Delta) S_{R}(p),
\end{equation}
With the help of this definition, the general Hurwitz partition function is written as
\begin{equation}
    Z_{g}(p^{(1)}, \dots, p^{(m)}| \beta) = \exp \left(\sum_{\Delta} \beta_{\Delta} \hat{W}_{\Delta} \right) Z_{g}(p^{(1)}, \dots, p^{(m)}| \beta = 0)
\end{equation}
The name ``cut-and-join" was given because of the property
\begin{equation}\label{cut-and-join_property}
    \hat{W}_{\Delta} \left( \frac{p_{\Delta'}}{z_{\Delta'}} \right) = \sum_{\Delta''} c_{\Delta \Delta'}^{\Delta''} \left( \frac{p_{\Delta''}}{z_{\Delta''}} \right)
\end{equation}
where $c_{\Delta \Delta'}^{\Delta''}$ are structure constants of the center of the group algebra of symmetric group ($ \Delta \circ \Delta' = c_{\Delta \Delta'}^{\Delta''} \Delta'' $).

Now we construct a simple way to generate the operators $\hat{W}_{\Delta}$: we use their representation via the Casimir operators $\hat{C}_{k}$. This can be done as follows. One should express the eigenvalues of the cut-and-join operators via the eigenvalues of the Casimir operators \cite[formula (53)]{mironov2020cut}:
\begin{equation}
    \phi_{R}(\Delta) = \sum_{\mu \vdash |\Delta|} S_{\mu}^{*}(p_{k}^{*} = C_k(R)) \frac{\psi_{\mu}(\Delta)}{z_{\Delta}},
\end{equation}
where $S_{\mu}^{*}(p^{*})$ denotes the shifted Schur polynomial (see below), and the eigenvalues of the Casimir operators are chosen as
\begin{equation}\label{gl_Casimirs}
    C_{k}(R) = \sum_{i=1}^{l(R)} \Big( (R_{i}-i)^{k} - (-i)^{k} \Big)
\end{equation}
Since both $\hat{W}_{\Delta}$ and $\hat{C}_{k}$ have the Schur polynomials as their eigenfunctions, then the cut-and-join operators may be expressed as
\begin{equation}\label{W_via_Casimirs_ordinary}
    \boxed{
    \hat{W}_{\Delta} = \sum_{\mu \vdash |\Delta|} S_{\mu}^{*}(p_{k}^{*} = \hat{C}_{k}) \frac{\psi_{\mu}(\Delta)}{z_{\Delta}}
    }
\end{equation}
If one knows explicit expressions for the Casimir operators from some other considerations, then, using this formula, one can write explicit expressions for the cut-and-join operators. In the rest of this section, we evaluate the first few examples of \eqref{W_via_Casimirs_ordinary} using the first few shifted Schur polynomials of the previous subsection, and check this formula directly.

\subsection{Ordinary cut-and-join operators via Casimir operators}

Now we use formula \eqref{W_via_Casimirs_ordinary} and calculations of the shifted Schur polynomials in order to explicitly write the first few $W$-operators:
\begin{equation}\label{first_few_examples}
    \begin{gathered}
    \hat{W}_{[1]} = \hat{C}_{1}\\
    \hat{W}_{[2]} = \frac{1}{2}\hat{C}_{2} + \frac{1}{2}\hat{C}_{1}\\
    \hat{W}_{[1,1]} = \frac{1}{2}\hat{C}_{1}^{2} - \frac{1}{2}\hat{C}_{1}\\
    \hat{W}_{[3]} = \frac{1}{3}\hat{C}_{3} + \frac{1}{2}\hat{C}_{2} - \frac{1}{2}\hat{C}_{1}^{2} + \frac{2}{3}\hat{C}_{1}\\
    \hat{W}_{[2,1]} = \frac{1}{2}\hat{C}_{2}\hat{C}_{1} - \hat{C}_{2} + \frac{1}{2}\hat{C}_{1}^{2} - \hat{C}_{1}\\
    \hat{W}_{[1,1,1]} = \frac{1}{6}\hat{C}_{1}^{3} - \frac{1}{2}\hat{C}_{1}^{2} + \frac{1}{3}\hat{C}_{1}
    \end{gathered}
\end{equation}
Now, using explicit expressions for the Casimir operators, one writes down the cut-and-join operators with the help of these formulas.

\textit{Remark 1.} One may obtain these expressions without using the shifted Schur polynomials but instead solving a linear system of equations expressing $\phi_{R}(\Delta)$ via $C_{k}(R)$ in the following way:
\begin{itemize}
    \item calculate all $\phi_{R}(\Delta)$ with the help of \eqref{phi_normalized_extended},

    \item calculate all $C_{\Delta}(R) = \prod_{k=1}^{l(\Delta)} C_{\Delta_{k}}(R)$,

    \item construct a system of linear equations
    \begin{equation}
        \phi_{R}(\Delta) = a_{R,1} C_{[|\Delta|]}(R) + a_{R,2} C_{[|\Delta|-1,1]}(R) + \dots + a_{R,N} C_{[1]}(R)
    \end{equation}
    for all $R$: $|R| \le |\Delta|$. There are exactly $N$ linear equations, therefore, the system of equations can be solved. One may check that the obtained expressions for $\phi_{R}(\Delta)$, and consequently for $\hat{W}_{\Delta}$, coincide with \eqref{first_few_examples}.
\end{itemize}

\vspace{0.5cm}
\textit{Remark 2.} Note that we use the Casimir eigenvalues \eqref{gl_Casimirs} that differ from another common choice,
\begin{equation}
    \widetilde{C}_{k}(R) = \sum_{i=1}^{l(R)} \left\{ \left(R_{i} - i + \frac{1}{2}\right)^{k} - \left(-i + \frac{1}{2}\right)^{k} \right\}
\end{equation}
The $W$-operators are expressed via these Casimir operators in a different way:
\begin{equation}
    \begin{gathered}
    \hat{W}_{[1]} = \hat{\widetilde{C}}_{1}\\
    \hat{W}_{[2]} = \frac{1}{2}\hat{\widetilde{C}}_{2}\\
    \hat{W}_{[1,1]} = \frac{1}{2}\hat{\widetilde{C}}_{1}^{2} - \frac{1}{2}\hat{\widetilde{C}}_{1}\\
    \end{gathered}
\end{equation}

\section{Spin Hurwitz numbers}

\subsection{Basic objects}

In the spin case, one needs to literally substitute every concept from the ordinary case to its spin counterpart. Let us list them in order.

\paragraph{1.} Counterparts of the ordinary Schur polynomials are the Q-Schur polynomials. To define them, one can start from the generating function for polynomials $P_{n,m}$:
\begin{equation}\label{Qij}
    \sum_{n,m = 0}^{\infty} P_{n,m} z_{1}^{n} z_{2}^{m} = \left\{ \exp\left( 2\sum_{k=0}^{\infty} \frac{p_{2k+1}}{2k+1} (z_{1}^{2k+1} + z_{2}^{2k+1}) \right) - 1 \right\} \frac{z_{1} - z_{2}}{z_{1} + z_{2}}
\end{equation}
From this definition, it is obvious that $P_{n,m}$ are antisymmetric. Define the antisymmetric matrix
\begin{equation}
    \big( M_{R}(p) \big)_{i,j} = P_{R_{i}, R_{j}}(p)
\end{equation}
for even $l(R)$ and if $l(R)$ is odd, add exactly one line $ P_{0,R_{j}}(p) $ (as if one adds a zero length line to the Young diagram $R$). Finally, the Q-Schur polynomials are defined as
\begin{equation}
    Q_{R}(p) := 2^{-l(R)/2} \Pf \Big( M_{R}(p) \Big) \equiv 2^{-l(R)/2} \sqrt{\det M_{R}(p)}
\end{equation}
This definition implies that the Q-Schur polynomials are equal to zero when $R$ contains two lines of coinciding lengths. The diagrams that do not have lines of equal length are called \textit{strict} partitions (SP).
%The normalization factor is convenient for the norm  of Q-Schur functions to be equal to 1.

\paragraph{2.} Counterparts of the symmetric group characters are characters of the Sergeev group $ \Psi_{R}(\Delta) $. The Q-Schur polynomials are connected with characters of the Sergeev group via the counterpart of \eqref{Schur_via_characters}:
\begin{equation}\label{Q_frobenius_formula}
    Q_{R}(p) = \sum_{\Delta \in \OP} \frac{\Psi_{R}(\Delta)}{z_{\Delta}} p_{\Delta}
\end{equation}
where OP denotes the \textit{odd} partitions, i.e. partitions with all parts odd. $ \Psi_{R}(\Delta) $ is equal to zero for $ R \notin \SP $ and provides an isomorphism between sets $\OP $ and $\SP$ (they have equal number of elements). Using \eqref{Q_frobenius_formula}, one can evaluate $ \Psi_{R}(\Delta) $ for all $ |R| = |\Delta| $.

The next step is to define properly normalized characters $\Phi_{R}(\Delta)$ like the ordinary case:
\begin{itemize}
    \item $ \mathfrak{d}_{R} := \frac{1}{2} Q_{R}(\delta_{k,1}) = \frac{\Psi_{R}([1^{|R|}])}{2|R|!} $

    $ \Phi_{R}(\Delta) := \frac{\Psi_{R}(\Delta)}{z_{\Delta} \mathfrak{d}_{R}} $; this definition is for $ |R| = |\Delta| $

The Q-Schur polynomials are then
    \begin{equation}\label{QSchur_via_normalized_characters}
        Q_{R}(p) = \sum_{\Delta \in \OP} \mathfrak{d}_{R} \Phi_{R}(\Delta) p_{\Delta}
    \end{equation}

    \item The definition extended to arbitrary $R$ and $\Delta$ is
    \begin{equation}\label{Qphi_normalized_extended}
        \Phi_{R}(\Delta) := \begin{cases}
        0, & |\Delta| > |R| \\
        \frac{(|R| - |\Delta| + r)!}{r! (|R| - |\Delta|)!} \Phi_{R}([\Delta, 1^{|R|-|\Delta|}]), & |\Delta| \le |R|
        \end{cases}
    \end{equation}
    where $r$ is the number of unit length lines in $\Delta$.
\end{itemize}

To be maximally explicit, we list the first few examples of $\Phi_{R}(\Delta)$:
\begin{equation}\label{Phi_table}
\begin{array}{cccccccccccc}
 R \backslash \Delta & [1] & [2] & [1,1] & [3] & [2,1] & [1,1,1] & [4] & [3,1] & [2,2] & [2,1,1] & [1,1,1,1] \\
 \left[1\right] & 2 & 0 & 0 & 0 & 0 & 0 & 0 & 0 & 0 & 0 & 0 \\
 \left[2\right] & 4 & 0 & 2 & 0 & 0 & 0 & 0 & 0 & 0 & 0 & 0 \\
 \left[1,1\right] & 0 & 0 & 0 & 0 & 0 & 0 & 0 & 0 & 0 & 0 & 0 \\
 \left[3\right] & 6 & 0 & 6 & 1 & 0 & 2 & 0 & 0 & 0 & 0 & 0 \\
 \left[2,1\right] & 6 & 0 & 6 & -2 & 0 & 2 & 0 & 0 & 0 & 0 & 0 \\
 \left[1,1,1\right] & 0 & 0 & 0 & 0 & 0 & 0 & 0 & 0 & 0 & 0 & 0 \\
 \left[4\right] & 8 & 0 & 12 & 4 & 0 & 8 & 0 & 4 & 0 & 0 & 2 \\
 \left[3,1\right] & 8 & 0 & 12 & -2 & 0 & 8 & 0 & -2 & 0 & 0 & 2 \\
 \left[2,2\right] & 0 & 0 & 0 & 0 & 0 & 0 & 0 & 0 & 0 & 0 & 0 \\
 \left[2,1,1\right] & 0 & 0 & 0 & 0 & 0 & 0 & 0 & 0 & 0 & 0 & 0 \\
 \left[1,1,1,1\right] & 0 & 0 & 0 & 0 & 0 & 0 & 0 & 0 & 0 & 0 & 0 \\
\end{array}
\end{equation}

\paragraph{3.} The spin Hurwitz numbers count ramified coverings with spin structures. A counterpart of the Fr\"obenius formula \eqref{Hurwitz_definition} in the spin case includes characters of the Sergeev group \cite{eskin2008theta, gunningham2016spin}. We use it as a definition of the spin Hurwitz numbers with the following normalization:
\begin{equation}\label{spin_Hurwitz_definition}
    \Hur^{(\spin)}_{g,n,p(\Sigma)}(\Delta_{1}, \Delta_{2}, \dots, \Delta_{m}) := \sum_{\substack{R \vdash n\\ R \in \SP}} (-1)^{p(\Sigma) \cdot l(R)} \mathfrak{d}_{R}^{2-2g} \Phi_{R}(\Delta_{1}) \Phi_{R}(\Delta_{2}) \dots \Phi_{R}(\Delta_{m})
\end{equation}
where $p(\Sigma) \in \mathbb{Z} / 2\mathbb{Z}$ and depends on the spin structure of the base surface $\Sigma$ (for more details, see \cite{eskin2008theta, gunningham2016spin, lee2012recursion, lee2018note, mironov2020cut}). We compare different normalization factors in the definition of the spin Hurwitz numbers used in the literature in Appendix A. The definition implies that the spin Hurwitz numbers exist only if all partitions $\Delta_{i}$ are odd partitions, and all $|\Delta_{i}| = n$. Similarly to the ordinary case, we construct the generating function with fixed number of ramification points
\begin{equation}\label{spin_Hurwitz_partition_easy}
    \sum_{n} \sum_{\Delta_{1}, \dots, \Delta_{m}} \Hur_{g,n,p(\Sigma)}^{(\spin)}(\Delta_{1}, \dots, \Delta_{m}) p_{\Delta_{1}}^{(1)} \dots p_{\Delta_{m}}^{(m)} = \sum_{R \in \SP} (-1)^{p(\Sigma) \cdot l(R)} \mathfrak{d}_{R}^{2-2g-m} Q_{R}(p^{(1)}) \dots Q_{R}(p^{(m)})
\end{equation}
which is a consequence of \eqref{QSchur_via_normalized_characters} and \eqref{spin_Hurwitz_definition} and looks similar to \eqref{Hurwitz_partition_function_easy}.

The general form of spin Hurwitz partition function that is a counterpart of \eqref{Hurwitz_partition_function_general} is
\begin{equation}\label{spin_Hurwitz_partition_general}
    Z_{g,p(\Sigma)}^{(\spin)} (p^{(1)}, \dots, p^{(m)}| \beta) = \sum_{R \in \SP}  (-1)^{p(\Sigma) \cdot l(R)} \left( e^{\sum_{\Delta} \beta_{\Delta} \Phi_{R}(\Delta)} \right) \mathfrak{d}_{R}^{2-2g-m} Q_{R}(p^{(1)}) \dots Q_{R}(p^{(m)})
\end{equation}
where $\beta_{\Delta}$ are again arbitrary parameters, and $\Phi_{R}(\Delta)$ are of the form \eqref{Qphi_normalized_extended}. Simple generating function \eqref{spin_Hurwitz_partition_easy} is obtained in the limit of $\beta_{\Delta} = 0$. Some generating functions in the spin case are known to be $\tau$-functions of the BKP hierarchy. Integrability again depends on the combination of $\beta_{\Delta}$ so that the Sergeev characters combine into a linear combination of certain Casimir eigenvalues defined in \eqref{Q_Casimirs} \cite{orlov2003hypergeometric, mironov2020cut}. The first non-trivial example of \eqref{spin_Hurwitz_partition_general} being a BKP $\tau$-function includes the combination of $\Phi_{R}([3])$ and $\Phi_{R}([1,1])$:
\begin{equation}
    Z^{(\spin)} (p, \bar{p}) = \sum_{R \in \SP} \left( e^{ u [\Phi_{R}([3]) + \frac{1}{2} \Phi_{R}([1,1])]} \right) Q_{R} \left( \frac{p}{2} \right) Q_{R} \left( \frac{\bar{p}}{2} \right)
\end{equation}
Note that it becomes a BKP $\tau$-function in variables $\frac{p_{k}}{2}$. Here $m=2, g=0, p(\Sigma \equiv S^{2})=0$, and parameters $ \beta_{[3]}=u, \beta_{[1,1]}=\frac{u}{2}, \beta_{\Delta} = 0 \; (\Delta \ne [3],[1,1])$.

\subsection{Shifted Q-Schur polynomials}

Similarly to the case of ordinary Hurwitz numbers, we need a notion of shifted Q-Schur polynomials. They were defined in \cite{ivanov2005interpolation}, for some their properties see \cite{mironov2020cut,orlov2021notes}. All the formulas for the shifted Schur polynomials are extended to the shifted Q-Schur polynomials with the change of determinant formulas to the Pfaffian ones. First of all, the counterpart of \eqref{shift_Schur_two_dets} is
\begin{equation}
    Q_{\mu}^{*}(x) := 2^{l(\mu)/2}\frac{\Pf (A_{\mu}(x))}{\Pf (A_{0}(x))}
\end{equation}
where we use the following notation. Let $n \ge l(\mu)$, $(x|n)$ be the same as in \eqref{shift_power_x}, $A_{0}(x)$ is the skew-symmetric $n \times n$ matrix:
\begin{equation}\label{Azero}
    (A_{0}(x_{1}, \dots, x_{n}))_{ij} = \frac{x_{i} - x_{j}}{x_{i} + x_{j}}
\end{equation}
for even $n$. For odd $n$ the matrix is $A_{0}(x_{1}, \dots, x_{n}, 0)$. In other words, in that case, it is of the size $(n+1) \times (n+1)$ defined as in \eqref{Azero} with the last variable $x_{n+1}$ equal to zero. In its turn, $A_{\mu}(x)$ is the skew-symmetric $(n+l(\mu)) \times (n+l(\mu))$ matrix:
\begin{equation}\label{Amu}
    A_{\mu}(x) = \begin{pmatrix}
    A_{0}(x) & B_{\mu}(x) \\
    -B_{\mu}^{T}(x) & 0
    \end{pmatrix}
\end{equation}
for even $(n+l(\mu))$, where $B_{\mu}(x)$ is the $n \times l(\mu)$ matrix:
\begin{equation}\label{Bmu}
    (B_{\mu}(x_{1}, \dots, x_{n}))_{ij} = (x_{i} | \mu_{l(\mu)+1-j})
\end{equation}
For odd $n+l(\mu)$ the matrix is $A_{\mu}(x_{1}, \dots, x_{n},0)$. It is of size $(n+l(\mu)+1) \times (n+l(\mu)+1)$ defined as in \eqref{Amu}, \eqref{Bmu} with the last variable $x_{n+1}$ equal to zero. In variance with the shifted Schur polynomials, the $p$-variables for the shifted Q-Schur polynomials are just the usual power sums
\begin{equation}
    p_{k} = \sum_{i\ge1} x_{i}^{k}
\end{equation}

The next step is to get a counterpart of the determinant formula \eqref{shift_Schur_det}. Let us consider a different expansion of the generating function for $P_{i,j}$ at the r.h.s. of (\ref{Qij}):
\begin{equation}
        \left\{ \exp\left( 2\sum_{k=0}^{\infty} \frac{p_{2k+1}}{2k+1} (z_{1}^{2k+1} + z_{2}^{2k+1}) \right) - 1 \right\} \frac{z_{1} - z_{2}}{z_{1} + z_{2}}={1\over z_1z_2} \sum_{k,l \ge 0} \frac{P_{k,l}^{*}}{(z_1^{-1}|k+1)(z_2^{-1}|l+1)}
        %=\sum_{n,m = 0}^{\infty} Q_{n,m} z_{1}^{n} z_{2}^{m}
\end{equation}
i.e. we make re-expansion
\begin{equation}
{1\over z_1z_2} \sum_{k,l \ge 0} \frac{P_{k,l}^{*}}{(z_1^{-1}|k+1)(z_2^{-1}|l+1)}
        =\sum_{n,m = 0}^{\infty} P_{n,m} z_{1}^{n} z_{2}^{m}
\end{equation}
with the l.h.s. understood as the Taylor expansion in $z_1,\ z_2$.
The first few terms of this re-expansion are
\begin{equation}
\begin{gathered}
    P_{0,0}^{*}=P_{0,0},\ \ \ \ \ P_{1,0}^*=P_{1,0},\ \ \ \ \ P_{0,1}^{*}=P_{0,1},\ \ \ \ \ P_{2,0}^{*}=P_{2,0}-P_{1,0},\ \ \ \ \ P_{1,1}^{*}=P_{1,1},\ \ \ \ \ P_{0,2}^{*}=P_{0,2}-P_{0,1}\\
    P_{2,1}^{*}=P_{2,1}-P_{1,1},\ \ \ \ \ P_{1,2}^{*}=P_{1,2}-P_{1,1},\ \ \ \ \ P_{3,0}^{*}=P_{3,0}-3P_{2,0}+2P_{1,0},\ \ \ \ \
P_{0,3}^{*}=P_{0,3}-3P_{0,2}+2P_{0,1}\\
P_{2,2}^{*}=P_{2,2}+P_{1,1}-P_{1,2}-P_{2,1},\ \ \ \ \ \hbox{etc.}
    \end{gathered}
\end{equation}
Thus, all $P_{k,l}^{*}$ can be expressed as linear combinations of $P_{i,j}$, and the expression for the shifted Q-Schur polynomials is given by the same Pfaffian formula as before but with deformed entries:
\begin{equation}
    Q_{\mu}^{*} = 2^{-l(\mu)/2} \Pf(P_{\mu_k,\mu_l}^{*})
\end{equation}
Here, again, if $l(\mu)$ is odd, one has to add the zero-length line to the Young diagram $\mu$.

Finally, a counterpart of the lower-triangular property \eqref{shifted_Schur_property}, \eqref{shifted_system} gives a straightforward way to generate the shifted Q-Schur polynomials in the $p$-variables:
\begin{equation}
    \begin{gathered}
    Q_{\mu}^{*}(p) = Q_{\mu}(p) + \sum_{\lambda \in \SP: |\lambda| < |\mu|} C_{\mu,\lambda} Q_{\lambda}(p)\\
    Q_{\mu}^{*}(x_{i} = R_{i}) = 0, \;\;\; \mu \nsubseteq R
    \end{gathered}
\end{equation}
If one uses this property as a definition, it completely determines the coefficients $C_{\mu,\lambda}$.

The first few examples of the shifted Q-Schur polynomials:
\begin{equation}
    \begin{gathered}
    Q_{[1]}^{*}(p) = \underbrace{\sqrt{2}p_{1}}_{=Q_{[1]}(p)}\\
    Q_{[2]}^{*}(p) = \underbrace{\sqrt{2} p_{1}^{2}}_{=Q_{[2]}(p)} - \sqrt{2}p_{1}\\
    Q_{[3]}^{*}(p) = \underbrace{ \frac{2\sqrt{2}}{3} p_{1}^{3} + \frac{\sqrt{2}}{3} p_{3}}_{=Q_{[3]}(p)} - 3\sqrt{2}p_{1}^{2} + 2\sqrt{2}p_{1}\\
    Q_{[2,1]}^{*}(p) = \underbrace{ \frac{2}{3}p_{1}^{3} - \frac{2}{3}p_{3} }_{=Q_{[2,1]}(p)}
    \end{gathered}
\end{equation}

\subsection{Spin cut-and-join operators}

Spin cut-and-join operators are defined similarly to their ordinary counterparts:
\begin{equation}\label{QW_definition}
    \hat{\mathcal{W}}_{\Delta} Q_{R}(p) := \Phi_{R}(\Delta) Q_{R}(p)
\end{equation}
but, in the spin case, $\Delta \in \OP$ and $R \in \SP$. General spin Hurwitz partition function can be written with the help of the spin cut-and-join operators:
\begin{equation}
    Z_{g}^{(\spin)}(p^{(1)}, \dots, p^{(m)}| \beta) = \exp \left(\sum_{\Delta} \beta_{\Delta} \hat{\mathcal{W}}_{\Delta} \right) Z_{g}^{(\spin)}(p^{(1)}, \dots, p^{(m)}| \beta = 0)
\end{equation}
The cut-and-join property for the spin $\hat{\mathcal{W}}$-operators similar to \eqref{cut-and-join_property} deserves further investigation.

As in the ordinary case, the spin cut-and-join operators $\hat{\mathcal{W}}_{\Delta}$ can be expressed via suitable Casimir operators $\hat{\mathcal{C}}_{k}$ with the Q-Schur polynomials as their eigenfunctions. Again, we express eigenvalues of the spin cut-and-join operators via eigenvalues of the Casimir operators \cite[formula (111)]{mironov2020cut} (an additional factor of 2 is necessary for $\Phi_{R}(\Delta)$ to be exactly \eqref{Phi_table}):
\begin{equation}\label{PhiQ}
    \Phi_{R}(\Delta) = 2\sum_{\mu \vdash |\Delta|} 2^{-|\mu|} Q_{\mu}^{*}(p_{k} = \mathcal{C}_{k}(R)) \frac{\Psi_{\mu}(\Delta)}{z_{\Delta}}
\end{equation}
where eigenvalues of the Casimir operators are chosen as
\begin{equation}\label{Q_Casimirs}
    \mathcal{C}_{k}(R) = \sum_{i=1}^{l(R)} R_{i}^{k}
\end{equation}
and the shifted Q-Schur functions are defined in the previous subsection. Expression for $\hat{\mathcal{W}}_{\Delta}$, which is a counterpart of \eqref{W_via_Casimirs_ordinary} follows from (\ref{PhiQ}):
\begin{equation}\label{QW_via_Casimirs_spin}
    \boxed{
    \hat{\mathcal{W}}_{\Delta} = 2\sum_{\mu \vdash |\Delta|} 2^{-|\mu|} Q_{\mu}^{*}(p_{k} = \hat{\mathcal{C}}_{k}) \frac{\Psi_{\mu}(\Delta)}{z_{\Delta}}
    }
\end{equation}
where $\hat{\mathcal{C}}_{k}$ are the Casimir operators with the eigenvalues (\ref{Q_Casimirs}), the Casimir operators being explicitly constructed in the next section.
This is a direct formula to evaluate the spin cut-and join operators with the help of Casimir operators.  In section 4, we calculate the Casimir operators and the cut-and-join operators in the spin case. Now we evaluate the first few examples of \eqref{QW_via_Casimirs_spin}.

\subsection{Spin cut-and-join operators via Casimir operators}
%Eigenvalues of Casimir operators in the spin case:

%We are going to express $ \Phi_{R}(\Delta) $ via these Casimir eigenvalues directly by solving system of linear equations. It was explained in remark 1. The only difference is that we omit zero lines $ R \notin SP $ and zero columns $ \Delta \notin OP $. We introduce the results of calculation that coincide with \cite[formula (102)]{mironov2020cut}:
%\begin{equation}
%    \begin{gathered}
%    \Phi_{R}([1]) = 2 \mathcal{C}_{1}(R) \\
%    \Phi_{R}([1,1]) = \mathcal{C}_{1}(R)^{2} - \mathcal{C}_{1}(R) \\
%    \Phi_{R}([3]) = \frac{1}{6} \mathcal{C}_{3}(R) - \frac{1}{2} \mathcal{C}_{1}(R)^{2} + \frac{1}{3} \mathcal{C}_{1}(R) \\
%    \Phi_{R}([1,1,1]) = \frac{1}{3} \mathcal{C}_{1}(R)^{3} - \mathcal{C}_{1}(R)^{2} + \frac{2}{3} \mathcal{C}_{1}(R)
%    \end{gathered}
%\end{equation}

Explicit calculating the shifted Q-Schur polynomials gives rise to manifest expressions for $\hat{\mathcal{W}}_{\Delta}$ in terms of the Casimir operators in the spin case using \eqref{QW_via_Casimirs_spin}:
\begin{equation}\label{first_few_QW_via_casimirs}
    \begin{gathered}
    \hat{\mathcal{W}}_{[1]} = 2 \hat{\mathcal{C}}_{1} \\
    \hat{\mathcal{W}}_{[1,1]} = \hat{\mathcal{C}}_{1}^{2} - \hat{\mathcal{C}}_{1} \\
    \hat{\mathcal{W}}_{[3]} = \frac{1}{6} \hat{\mathcal{C}}_{3} - \frac{1}{2} \hat{\mathcal{C}}_{1}^{2} + \frac{1}{3} \hat{\mathcal{C}}_{1} \\
    \hat{\mathcal{W}}_{[1,1,1]} = \frac{1}{3} \hat{\mathcal{C}}_{1}^{3} - \hat{\mathcal{C}}_{1}^{2} + \frac{2}{3} \hat{\mathcal{C}}_{1}
    \end{gathered}
\end{equation}
The expression for $ \hat{\mathcal{W}}_{[3]} $ was found in \cite[section 5.2]{mironov2020around}.

\section{Explicit expression for spin $\mathcal{W}$-operators}

In this section, we follow \cite{mironov2020around} to calculate all the necessary Casimir operators using bosonization formulas. It allows us to construct an explicit expression for the desired spin cut-and-join operators. Let us define the vertex operators
\begin{equation}
    \hat{V}(z, \textbf{p}) = \frac{1}{\sqrt{2}} \exp\left( \sum_{m \in \mathbb{Z}_{odd}^{+} } \frac{2}{m} z^{m} p_{m} \right) \exp\left( \sum_{m \in \mathbb{Z}_{odd}^{+} } -z^{-m} \frac{\partial}{\partial p_{m}} \right)
\end{equation}
and use their product to construct a generating function of operators $\hat{\Omega}_{m}^{(n)}$:
\begin{equation}
    \frac{1}{2} \hat{V}(z e^{y/2}, \textbf{p}) \hat{V}(-z e^{-y/2}, \textbf{p}) - \frac{1}{4} \frac{1 + e^{-y}}{1 - e^{-y}} = \sum_{n=0}^{\infty} \frac{1}{n!} y^{n}  \sum_{m \in \mathbb{Z}} z^{m} \hat{\Omega}_{m}^{(n)}(\textbf{p})
\end{equation}
The operators $ \hat{\Omega}_{m}^{(n)}(\textbf{p}) $ generate the $ W_{1+\infty}^{B} $ algebra. The $n =0$ set of operators generates a  Heisenberg subalgebra, the $ n = 1$ set, a Virasoro subalgebra. However, we are interested in the commutative subalgebra of zero modes $ \hat{\Omega}_{0}^{(n)} $ for odd $n$:
\begin{equation}
    [ \hat{\Omega}_{0}^{(n)}, \hat{\Omega}_{0}^{(m)} ] = 0
\end{equation}
The first few examples are:
\begin{equation}
    \begin{gathered}
    \hat{\Omega}_{0}^{(1)} = \sum_{n \in \mathbb{Z}_{odd}^{+}} n p_{n} \frac{\partial}{\partial p_{n}} \\
    \hat{\Omega}_{0}^{(3)} = \frac{1}{2} \sum_{n \in \mathbb{Z}_{odd}^{+}} (n^{3} + n) p_{n} \frac{\partial}{\partial p_{n}} + 4 \sum_{n_{1}, n_{2}, n_{3} \in \mathbb{Z}_{odd}^{+}} (n_{1} + n_{2} + n_{3}) p_{n_{1}} p_{n_{2}} p_{n_{3}} \frac{\partial}{\partial p_{n_{1} + n_{2} + n_{3}}} + \\
    3 \sum_{n_{1} + n_{2} = n_{3} + n_{4} \in \mathbb{Z}_{odd}^{+}} n_{1} n_{2} p_{n_{3}} p_{n_{4}} \frac{\partial}{\partial p_{n_{1}}} \frac{\partial}{\partial p_{n_{2}}} + \sum_{n_{1}, n_{2}, n_{3} \in \mathbb{Z}_{odd}^{+}} n_{1}n_{2}n_{3} p_{n_{1} + n_{2} + n_{3}} \frac{\partial^{3}}{\partial p_{n_{1}} \partial p_{n_{2}} \partial p_{n_{3}}}
    \end{gathered}
\end{equation}
The operators $ \hat{\Omega}_{0}^{(n)} $ act on the Q-Schur polynomials as
\begin{equation}
    \hat{\Omega}_{0}^{(n)} Q_{R}(\textbf{p}) = \mathcal{C}_{n}(R) Q_{R}(\textbf{p})
\end{equation}
that is, the Q-Schur functions are their eigenfunctions with the eigenvalues equal to the Casimir eigenvalues in the spin case \eqref{Q_Casimirs}. Therefore, we conclude that $ \hat{\Omega}_{0}^{(n)} = \hat{\mathcal{C}}_{n} $. Now is it possible to write down an explicit expression for every $ \hat{\mathcal{W}}_{\Delta} $ in \eqref{first_few_QW_via_casimirs}:
\begin{equation}
    \hat{\mathcal{W}}_{[1]} = 2 \sum_{n \in \mathbb{Z}_{odd}^{+}} n p_{n}  \frac{\partial}{\partial p_{n}} = 2 \sum_{a=0}^{\infty} (2a+1) p_{2a+1} \frac{\partial}{\partial p_{2a+1}}
\end{equation}

\begin{align}
    \begin{split}
    \hat{\mathcal{W}}_{[1,1]} = \left( \sum_{n \in \mathbb{Z}_{odd}^{+}} n p_{n} \frac{\partial}{\partial p_{n}}  \right)^{2} - \sum_{n \in \mathbb{Z}_{odd}^{+}} n p_{n} \frac{\partial}{\partial p_{n}} = \sum_{n,m \in \mathbb{Z}_{odd}^{+}} n m p_{n} p_{m} \frac{\partial^{2}}{\partial p_{n} \partial p_{m}} + \sum_{n \in \mathbb{Z}_{odd}^{+}} (n^{2} - n) p_{n} \frac{\partial}{\partial p_{n}} = \\
    = \sum_{a=0}^{\infty} 2a (2a+1) p_{2a+1} \frac{\partial}{\partial p_{2a+1}} + \sum_{a,b=0}^{\infty} (2a+1) (2b+1) p_{2a+1} p_{2b+1} \frac{\partial^{2}}{\partial p_{2a+1} \partial p_{2b+1}}
    \end{split}
\end{align}

\begin{align}\label{cut_and_join_3}
    \begin{split}
    \hat{\mathcal{W}}_{[3]} = \frac{1}{12} \sum_{n \in \mathbb{Z}_{odd}^{+}} (n^{3} + n) p_{n} \frac{\partial}{\partial p_{n}} + \frac{2}{3} \sum_{n_{1}, n_{2}, n_{3} \in \mathbb{Z}_{odd}^{+}} (n_{1} + n_{2} + n_{3}) p_{n_{1}} p_{n_{2}} p_{n_{3}} \frac{\partial}{\partial p_{n_{1} + n_{2} + n_{3}}} + \\
    + \frac{1}{2} \sum_{n_{1} + n_{2} = n_{3} + n_{4} \in \mathbb{Z}_{odd}^{+}} n_{1} n_{2} p_{n_{3}} p_{n_{4}} \frac{\partial}{\partial p_{n_{1}}} \frac{\partial}{\partial p_{n_{2}}} + \frac{1}{6} \sum_{n_{1}, n_{2}, n_{3} \in \mathbb{Z}_{odd}^{+}} n_{1}n_{2}n_{3} p_{n_{1} + n_{2} + n_{3}} \frac{\partial^{3}}{\partial p_{n_{1}} \partial p_{n_{2}} \partial p_{n_{3}}}  - \\
    -\frac{1}{2} \left( \sum_{n \in \mathbb{Z}_{odd}^{+}} n p_{n} \frac{\partial}{\partial p_{n}}  \right)^{2} + \frac{1}{3} \sum_{n \in \mathbb{Z}_{odd}^{+}} n p_{n} \frac{\partial}{\partial p_{n}} = \\
    = \frac{2}{3} \sum_{n,m,k \in \mathbb{Z}_{odd}^{+}} (n+m+k) p_{n} p_{m} p_{k} \frac{\partial}{\partial p_{n+m+k}} + \frac{1}{6} \sum_{n,m,k \in \mathbb{Z}_{odd}^{+}} n m k p_{n+m+k} \frac{\partial^{3}}{\partial p_{n} \partial p_{m} \partial p_{k}} + \\
    + \frac{1}{2} \sum_{n+m=k+l \in \mathbb{Z}_{odd}^{+}} k l p_{n} p_{m} \frac{\partial^{2}}{\partial p_{k} \partial p_{l}} - \frac{1}{2} \sum_{n,m \in \mathbb{Z}_{odd}^{+}} n m p_{n} p_{m} \frac{\partial^{2}}{\partial p_{n} \partial p_{m}} + \sum_{n \in \mathbb{Z}_{odd}^{+}} \frac{n}{12} (n-1) (n-5) p_{n} \frac{\partial}{\partial p_{n}}
    \end{split}
\end{align}

\begin{align}
    \begin{split}
    \hat{\mathcal{W}}_{[1,1,1]} = \frac{1}{3} \left( \sum_{n \in \mathbb{Z}_{odd}^{+}} n p_{n} \frac{\partial}{\partial p_{n}}  \right)^{3} - \left( \sum_{n \in \mathbb{Z}_{odd}^{+}} n p_{n} \frac{\partial}{\partial p_{n}}  \right)^{2} + \frac{2}{3} \sum_{n \in \mathbb{Z}_{odd}^{+}} n p_{n} \frac{\partial}{\partial p_{n}} = \\
    = \sum_{n \in \mathbb{Z}_{odd}^{+}} \frac{n}{3} (n-1) (n-2) p_{n} \frac{\partial}{\partial p_{n}} + \sum_{n,m \in \mathbb{Z}_{odd}^{+}} n m (n-1) p_{n} p_{m} \frac{\partial^{2}}{\partial p_{n} \partial p_{m}} + \frac{1}{3} \sum_{n,m,k \in \mathbb{Z}_{odd}^{+}} n m k p_{n} p_{m} p_{k} \frac{\partial^{3}}{\partial p_{n} \partial p_{m} \partial p_{k}}
    \end{split}
\end{align}
Any concrete operator is obtained in the algorithmic way, and coincides with \cite[formulas (94), (95)]{mironov2020cut}, where such expressions were obtained by solving a system of equations in the space of operators with the restriction \eqref{QW_definition}. Note that in \cite[(101)]{mironov2020cut} there should be the "minus" sign in front of $\hat{\mathfrak{W}}_{[1,1]}$. At the first glance, it may seem that $\hat{\mathfrak{W}}_{[3]}$ does not contain the last term of \eqref{cut_and_join_3} with just one derivative, however they actually do coincide.

\section{Conclusion}

The ordinary cut-and-join operators have a large number of applications in the theory of Hurwitz partition functions. The major convenience of such operators is a possibility of their direct realization in a matrix form.  In the meanwhile, any matrix realization for their spin counterparts, generating the spin Hurwitz partition functions, is still unknown. However, the spin operators possess certain properties similar to those in the ordinary case that allow one to calculate them in an \textit{algorithmic} way without matrix realization. In this paper, we discussed these properties, and explained how to use them in order to calculate the spin cut-and-join operators in terms of derivatives in $p$-variables. This becomes possible because of a connection between the cut-and-join operators and specially chosen Casimir operators, which can be calculated directly as zero modes of the $W_{1+\infty}^{B}$ algebra. We explained this connection in the ordinary case, and then extended it to the spin case. An essential part of this connection is encoded by the shifted Schur and Q-Schur polynomials. Internal structure of the spin cut-and-join operators as well as their cut-and-join property deserve further investigation.

\section*{Acknowledgements}

%The work is partly??? supported by the ??? grant.
This work was funded by the Russian Science Foundation (Grant No.20-71-10073).

\bigskip

\section*{Appendix A. Normalization factors for spin Hurwitz numbers}

In this section, we compare different normalization factors in the spin Hurwitz numbers and provide the first few examples. We consider three normalization factors:
\begin{enumerate}
    \item[{\bf 1.}] In the present paper, the spin Hurwitz numbers are defined in \eqref{spin_Hurwitz_definition}:
\begin{equation}\label{one}
    \Hur^{(\spin)}_{g,n,p(\Sigma)}(\Delta_{1}, \Delta_{2}, \dots, \Delta_{m}) := \sum_{\substack{R \vdash n\\ R \in \SP}} (-1)^{p(\Sigma) \cdot l(R)} \mathfrak{d}_{R}^{2-2g} \Phi_{R}(\Delta_{1}) \Phi_{R}(\Delta_{2}) \dots \Phi_{R}(\Delta_{m})
\end{equation}
Their simple generating function defined in \eqref{spin_Hurwitz_partition_easy} is
\begin{equation}
    G_{1} = \sum_{R \in \SP} (-1)^{p(\Sigma) \cdot l(R)} \mathfrak{d}_{R}^{2-2g-m} Q_{R}(p^{(1)}) \dots Q_{R}(p^{(m)})
\end{equation}

\item[{\bf 2.}] The definition of the spin Hurwitz numbers in \cite[formula (91)]{mironov2020cut} differs by the power of 2:
\begin{equation}\label{two}
    \widetilde{\Hur}^{(\spin)}_{g,n,p(\Sigma)}(\Delta_{1}, \Delta_{2}, \dots, \Delta_{m}) := 2^{(n-2)(g-1)} \sum_{\substack{R \vdash n\\ R \in \SP}} (-1)^{p(\Sigma) \cdot l(R)} \mathfrak{d}_{R}^{2-2g} \Phi_{R}(\Delta_{1}) \Phi_{R}(\Delta_{2}) \dots \Phi_{R}(\Delta_{m})
\end{equation}
This normalization factor comes into each term of the corresponding generating function:
\begin{equation}
    G_{2} = \sum_{R \in \SP} 2^{(|R|-2)(g-1)} (-1)^{p(\Sigma) \cdot l(R)} \mathfrak{d}_{R}^{2-2g-m} Q_{R}(p^{(1)}) \dots Q_{R}(p^{(m)})
\end{equation}

\item[{\bf 3.}] One more normalization factor can be found in \cite[Theorem 1.10]{gunningham2016spin}. After rewriting it in our common notation,
\begin{equation}\label{three}
    H^{(\spin)}_{g,n,p(\Sigma)}(\Delta_{1}, \dots, \Delta_{m}) := 2^{\frac{1}{2}\left[ \sum_{i=1}^{m} (l(\Delta_{i}) - n) \right] - m} \cdot 2^{(n-2)(g-1)} \sum_{\substack{R \vdash n\\ R \in \SP}} (-1)^{p(\Sigma) \cdot l(R)} \mathfrak{d}_{R}^{2-2g} \Phi_{R}(\Delta_{1}) \dots \Phi_{R}(\Delta_{m})
\end{equation}
Note that the power of 2 in front of the sum is always an integer for non-zero Hurwitz number because of Riemann-Hurwitz formula connecting the ramification data and the Euler characteristic of the covering surface $\widetilde{\Sigma}$:
\begin{equation}
    \chi(\widetilde{\Sigma}) = 2 - 2\widetilde{g} = 2n(1-g) + \sum_{i=1}^{m} (l(\Delta_{i}) - n)
\end{equation}
Since this normalization factor depends on partitions $\Delta_{k}$, it changes the dependence on $p$-variables in the corresponding generating functions:
\begin{equation}
    G_{3} = \sum_{R \in \SP} 2^{M_{g,m}(|R|)} (-1)^{p \cdot l(R)} \mathfrak{d}_{R}^{2-2g-m} Q_{R}(\sqrt{2} p^{(1)}) \dots Q_{R}(\sqrt{2} p^{(m)})
\end{equation}
where $M_{g,m}(|R|) = -\frac{m|R|}{2} - m + (|R|-2)(g-1)$.
\end{enumerate}

Comparing definitions \eqref{one}, \eqref{two}, \eqref{three} of the spin Hurwitz numbers, one can see the following relations between them:
\begin{align}
    \begin{split}
    H_{g,n,p(\Sigma)}^{(\spin)}(\Delta_{1}, \dots, \Delta_{m}) = \\
    2^{\frac{1}{2}\left[ \sum_{i=1}^{m} (l(\Delta_{i}) - n) \right] - m} \cdot \widetilde{\Hur}_{g,n,p(\Sigma)}^{(\spin)}(\Delta_{1}, \dots, \Delta_{m}) = \\
    2^{\frac{1}{2}\left[ \sum_{i=1}^{m} (l(\Delta_{i}) - n) \right] - m} \cdot 2^{(n-2)(g-1)} \cdot \Hur_{g,n,p(\Sigma)}^{(\spin)}(\Delta_{1}, \dots, \Delta_{m})
    \end{split}
\end{align}

There are simple formulas for the first several non-trivial spin Hurwitz numbers \cite{lee2012recursion} in normalization of \eqref{two}:
\begin{equation}
    \begin{gathered}
    \widetilde{\Hur}_{g,3,p(\Sigma)}^{(\spin)}(\underbrace{[3],[3],\dots}_{k}) = 3^{2g-2} [(-1)^{k} 2^{k+g-1} + (-1)^{p(\Sigma)}] \\
    \widetilde{\Hur}_{g,4,p(\Sigma)}^{(\spin)}(\underbrace{[3,1],[3,1],\dots}_{k}) = (3!)^{2g-2} \cdot 2^{k} [(-1)^{p(\Sigma)} 2^{k+g-1} + (-1)^{k}]
    \end{gathered}
\end{equation}

%\bibliographystyle{ieeetr}
%\bibliography{references}

\end{document}